\documentclass[journal]{IEEEtran}
\pdfoutput=1
\usepackage{cite}
\usepackage{amsfonts}
\usepackage{color}
\usepackage[ruled,linesnumbered]{algorithm2e}

\ifCLASSINFOpdf
\fi

\usepackage{amsmath}

\usepackage{graphicx}
\usepackage{algorithmic}
\usepackage{threeparttable}
\usepackage{titletoc}
\usepackage{hyperref}

\usepackage{multirow}

\begin{document}
\title{Deep Covariance Alignment for Domain Adaptive Remote Sensing Image Segmentation
}

\author{Linshan Wu, Ming Lu, and Leyuan Fang, \IEEEmembership{Senior Member,~IEEE}%

\thanks{ This work was supported in part by the National Natural Science Fund of China under Grant 61922029, in part by the Science and Technology Plan Project Fund of Hunan Province under Grant 2019RS2016, and in part by the Key
Research and Development Project of Science and Technology Plan of Hunan Province under Grant 2021SK2039, and in Part by the Natural Science Fund of Hunan Province under Grant 2021JJ30003.(Corresponding author: Leyuan Fang.)}

\thanks{Linshan Wu is with the College of Electrical and Information Engineering, Hunan University, Changsha 410082, China, and also with the State Key Laboratory of Integrated Services Networks, Xidian University, Xian, 710126,
China.(e-mail: linshanwu@hnu.edu.cn).}
\thanks{Ming Lu is with the College of Electrical and Information Engineering, Hunan University, Changsha 410082, China(e-mail: 1148462196@qq.com).}
\thanks{Leyuan Fang is with the College of Electrical and Information Engineering, Hunan University, Changsha 410082, China, and also with the Peng Cheng Laboratory, Shenzhen 518000, China (e-mail: fangleyuan@gmail.com).}
}

\maketitle
\begin{abstract}
Unsupervised domain adaptive (UDA) image segmentation has recently gained increasing attention, aiming to improve the generalization capability for transferring knowledge from the source domain to the target domain. However, in high spatial resolution remote sensing image (RSI), the same category from different domains (\emph{e.g.}, urban and rural) can appear to be totally different with extremely inconsistent distributions, which heavily limits the UDA accuracy. To address this problem, in this paper, we propose a novel Deep Covariance Alignment (DCA) model for UDA RSI segmentation. The DCA can explicitly align category features to learn shared domain-invariant discriminative feature representations, which enhance the ability of model generalization. Specifically, a Category Feature Pooling (CFP) module is first employed to extract category features by combining the coarse outputs and the deep features. Then, we leverage a novel Covariance Regularization (CR) to enforce the intra-category features to be closer and the inter-category features to be further separate. Compared with the existing category alignment methods, our CR aims to regularize the correlation between different dimensions of the features, and thus performs more robustly when dealing with the divergent category features of imbalanced and inconsistent distributions. Finally, we propose a stagewise procedure to train the DCA in order to alleviate the error accumulation. Experiments on both Rural-to-Urban and Urban-to-Rural scenarios of the LoveDA dataset \cite{1} demonstrate the superiority of our proposed DCA over other state-of-the-art UDA segmentation methods. Code is available at \emph{\href{https://github.com/Luffy03/DCA}{https://github.com/Luffy03/DCA}}.
\end{abstract}


%
\IEEEpeerreviewmaketitle

\section{Introduction}

\IEEEPARstart{R}{emote} sensing image (RSI) segmentation aims at assigning the corresponding pixel-wise land-cover type at every image pixel, which plays an increasingly significant role for many applications \cite{2, 3, 4, 5}. However, the large requirement of labeled training samples and the diverse styles among geographic areas (\emph{e.g.}, urban and rural) heavily limited the development of RSI segmentation. One common solution for the lack and discrepancy of data is the unsupervised domain adaptation (UDA), whose goal is to improve the model generalizability from different domains. UDA aims to adapt the models trained from the labeled source domain to the unlabeled target domain, and thus alleviate the lack of annotated training samples, which has been attracting much attention in RSI segmentation task.

Recently, with the success of semantic segmentation methods \cite{34,35,36,37,38,42,44} based on deep CNNs \cite{30, 39}, UDA segmentation has been rapidly developed. In UDA segmentation, the recent works can be generally divided into two groups, \emph{i.e.}, adversarial training (AT) methods \cite{6,7,8,9} and self-training (ST) methods \cite{10,11,12}. The AT methods adopt a feature extractor to capture domain-invariant features and perform a discriminator to distinguish them. Tsai \emph{et.al} \cite{6} construct a multi-level adversarial network (AdaptSeg) to effectively perform output space domain adaptation at different feature levels. Luo \emph{et.al} \cite{7} design a category-level adversarial network (CLAN) to conduct a more delicate level domain calibration. Wang \emph{et.al} \cite{8} proposed a fine-grained adversarial learning framework (FADA) to align shared features. A transferable normalization (TransNorm) method in \cite{9} was further proposed to improve the transferability in UDA. However, these methods are almost based on GAN \cite{16} for adversarial training, which is difficult to train.

In addition, the ST methods have been also widely utilized in UDA segmentation task, which involves the model trained in the source domain to generate pseudo labels for the target domain and finally finetunes the model. Typically, Lian \emph{et.al} \cite{10} proposed PyCDA to construct self-motivated pyramid curriculums for UDA. Zou \emph{et.al} \cite{11} further proposed a class-balanced self-training (CBST) strategy to avoid the gradual dominance of large classes in pseudo-label generation, and introduce spatial priors to refine the generated pseudo-labels. An instance adaptive self-training (IAST) method is also proposed in \cite{12} to select balanced samples. The ST methods always work in a coarse-to-fine manner, which are usually trained with a stagewise mechanism.

The previous UDA methods have been also applied to RSI segmentation task \cite{17,18,19,20,21}. Most of them aim to preform photometric alignment by generative adversarial network (GAN) \cite{16}, which can be classified as AT methods. Despite some promising results of these methods have been achieved, actually, the advancements of these algorithms are limited for several reasons. First, as in the RSIs, the manifestation of the land-cover is always completely different. Particularly, the same category from diverse areas, \emph{i.e.}, urban and rural, can appear to be totally different in object scales and spectral values. As shown in Fig. 1 (a), buildings and roads from urban areas and rural areas are with extremely large discrepancy. Second, the imbalanced and inconsistent category distributions also pose a special challenge for the UDA RSI segmentation. As shown in Fig. 1 (b), the urban and rural areas have greatly different category distributions, which further increases the difficulty of model generalization in the UDA RSI segmentation task. 

\begin{figure}
	\centering
	\includegraphics[width=1\linewidth]{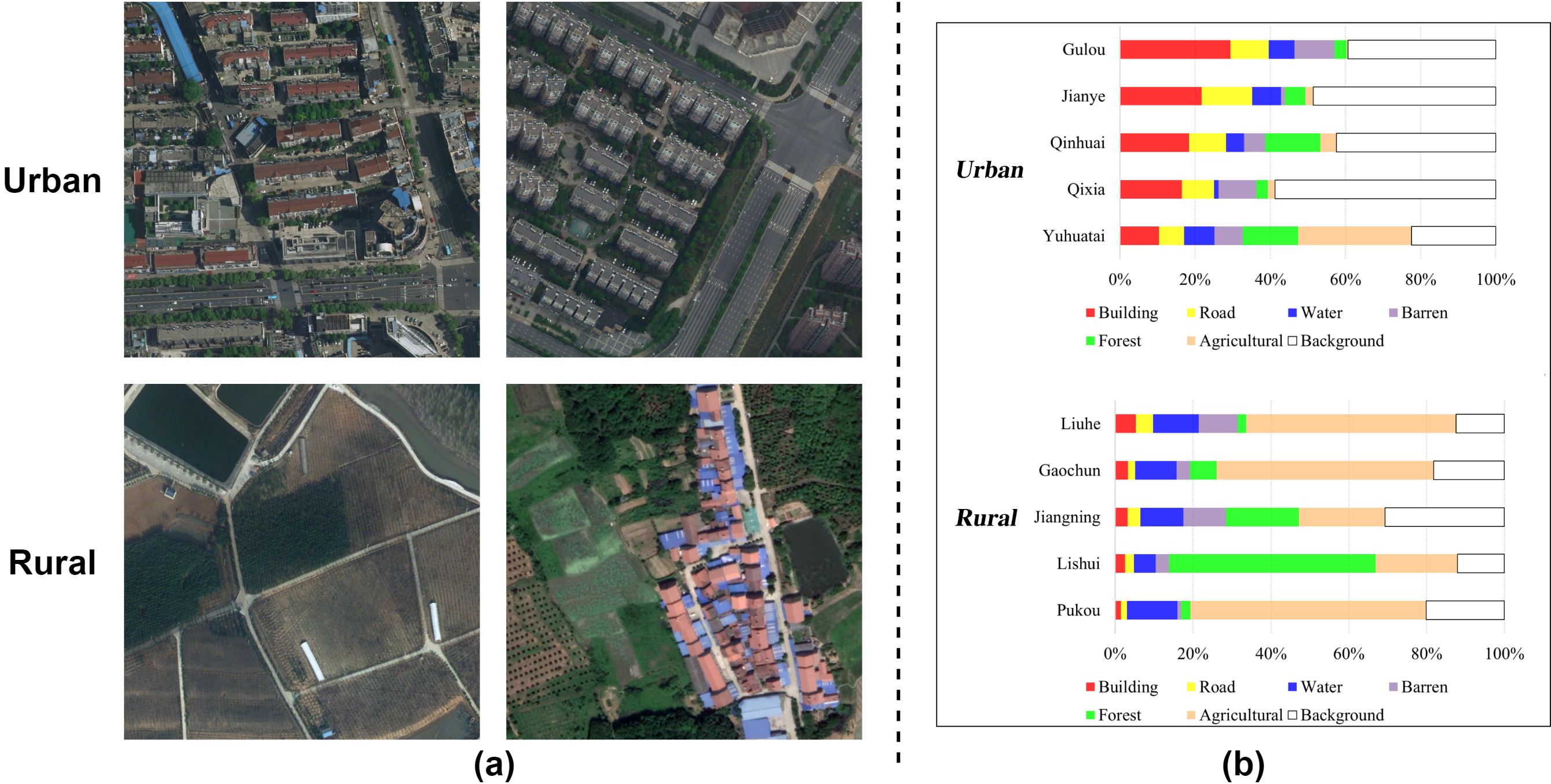}
	\caption{The difference between urban and rural scenes in LoveDA dataset [1]. (a) shows the discrepancy of categories between different domains, (b) shows the imbalanced and inconsistent category distributions. The details can be seen in \cite{1}.}
	\label{fig1}
\end{figure}

To tackle these issues, one direction is to align category-level features between two different domains. The works in \cite{13, 14} adopt category anchors computed on the source domain to guide the alignment between the two domains, which constrains the category centers with a simple Euclidean distance. The problem of this design is that it does not consider the relationship of inter-category features. Thus, triplet loss is further utilized in \cite{15} to regularize the distance between inter-category features. Although the above category alignment methods have produced some promising results, for UDA RSI segmentation task, there are still some urgent issues. First, for RSIs, the larger intra-category variance and the more imbalanced category distributions between two domains will heavily mislead the calculation of distance between different category features. In addition, it is very difficult to learn the decision boundary when adjusting the relative magnitude of intra-category and inter-category simultaneously, which always requires a complex handcrafted threshold setting.

To address the above issues, in this paper, we propose a novel Deep Covariance Alignment (DCA) method for UDA RSI segmentation. This model explicitly aligns category features to learn shared domain-invariant discriminative knowledge from the source domain to the target domain. First, a Category Feature Pooling (CFP) module is employed to extract category features by combining the coarse outputs and the deep features. Then, we leverage a novel Covariance Regularization (CR) to enforce the intra-category features to be closer and the inter-category features to be further separate. Compared with existing category alignment methods, our CR aims to regularize the correlation between different dimensions of the features, and thus perform more robustly when dealing with the divergent category features of imbalanced and inconsistent distributions. In addition, our CR can be trained without any other complex handcrafted settings. Finally, in order to alleviate the error accumulation, we propose a ST based stagewise training mechanism for our proposed DCA. 

The remainder of this paper is organized as follows: Section II reviews the related works. Section III describes the details of out proposed method. Section IV conducts the experiments to verify the effectiveness of our proposed method and comparisons with other methods. Finally, we conclude this paper and suggest some future works in Section V.

\section{Related Work}

\subsection{UDA RSI Segmentation}

Recently, UDA RSI segmentation has achieved great progress with several methods proposed \cite{17,18,19,20,21}. Most of these methods focus on preforming photometric alignment via generative adversarial network (GAN) \cite{16} to align the source images and the target images in the image space, feature space, and output space. Tasar \emph{et.al.} \cite{17} proposed color mapping generative adversarial networks (ColorMapGANs) to generate fake training images for fine-tuning the already trained classifiers. Ji \emph{et.al.} \cite{18} further use GAN to align multi-source RSIs. However, these AT based models are hard to train, and they generally align distributions from different domains, but do not actually obtain mappings between features from different domains. Other approaches adopt the idea of ST by generating pseudo-labels for samples in the target domain and providing extra supervision to the classifier \cite{22, 23}. For example, Tong \emph{et.al.} \cite{22} proposed a ST based method for UDA land-cover semantic segmentation, using a transferrable deep model. However, error-prone pseudo-labels will easily mislead the classifier and accumulate errors, which limit the effectiveness of the ST methods.

Despite some promising results have been achieved, most of these methods ignore the diverse styles among different geographic areas. While for urban and rural areas, in particular, the manifestation of the land cover is completely different in the class distributions, object scales, and pixel spectra. In order to improve the model generalizability for UDA RSI segmentation, Wang \emph{et.al.} \cite{1} originally created a groundbreaking LoveDA dataset, which contains diverse urban and rural RSIs. The dataset aims to promote the capacity of model generalization between the urban domain and the rural domain, which is seriously limited by the large category features discrepancy and inconsistent category distributions. 

\subsection{Category Alignment methods in UDA}
Although the fact that the domain gap can be minimized by GAN based AT methods mentioned above, there is no guarantee that features from different categories can be well separated. Thus, the idea of category-level feature alignment was also exploited in \cite{7,24} for UDA segmentation. Other approaches were proposed to match the local joint distributions of features and categories \cite{25,26}. Category labels were further introduced in \cite{27} to enforce global semantic constraints on the distribution of predicted labels. The ideas of minimizing the entropy (uncertainty) of the output \cite{28} have also been exploited to implicitly enforce category-level alignment.

In contrast to the implicit feature alignment in the aforementioned methods, some works \cite{13,14,15} propose to directly aligns category-wise features in both domains based on a category anchor-guided method, which have achieved more competitive performance. Among them, \cite{13,14} adopt category anchors computed on the source domain to guide the alignment between the two domains, which can be regarded as a hard constraint on the category centers. The hard constraint is a distance $d$ between different category features $f_1$ and $f_2$, which can be simply defined by a mean square error (MSE) function as
\begin{equation}\label{distance}
	d(f_1, f_2) = \parallel f_1 - f_2 \parallel
\end{equation}
where $\parallel  \parallel$ represents a Euclidean distance. $f_1$ and $f_2$ represent two features of the same category from two different domains, respectively. This distance $d$ is used to drive the intra-category closer by a loss function $L_{mse}$:
\begin{equation}\label{MSE}
	L_{mse}(f_1, f_2) = \frac{1}{N*N}\sum_{i}^{N} d(f^i_1, f^i_2)
\end{equation}
where $N$ represents the total number of categories, $f_1^i$ and  $f_2^i$ are the $i_{th}$ category feature from source domain and target domain, respectively.

However, this strategy does not explicitly enlarge the margins between the different category features centers. Thus, Ma \emph{et.al.} \cite{15} proposed a category-oriented triplet loss for the source domain that imposes a soft constraint to regularize category centers, actively making inter-category distances in a high-level feature space larger than intra-category distances by a specified margin. The triplet loss $L_{triplet}$ is formulated as follows:
\begin{equation}\label{Triplet}
\begin{aligned}
&L_{triplet}(f_1, f_2) \\ &= \frac{1}{N\times{N}}\sum_{i}^{N} \sum_{j}^{N} max(d(f^i_1, f^i_2)\big|_{i=j}-d(f^i_1, f^i_2)\big|_{i\neq j}+\alpha, 0)
\end{aligned}
\end{equation}
where ${\alpha}$ represents a prescribed margin. However, for different categories with large discrepancy, the best margins for training are also different, which requires a well-experienced handcrafted setting. Thus, work in \cite{15} only applies the $L_{triplet}$ to the source domain images, which have reliable category labels to supervise.

\section{Proposed DCA Method}
In this section, we introduce the architecture of our proposed DCA and the detailed training process. First, we introduce the category feature pooling (CFP) module to extract the category features. Then, we describe the formulation and motivations of our proposed Covariance Regularization (CR), and its application to align intra-domain and cross-domain features. Finally, we introduce how to train our proposed DCA.

\subsection{Category Feature Pooling (CFP) module}
\begin{figure}
	\centering
	\includegraphics[width=1\linewidth]{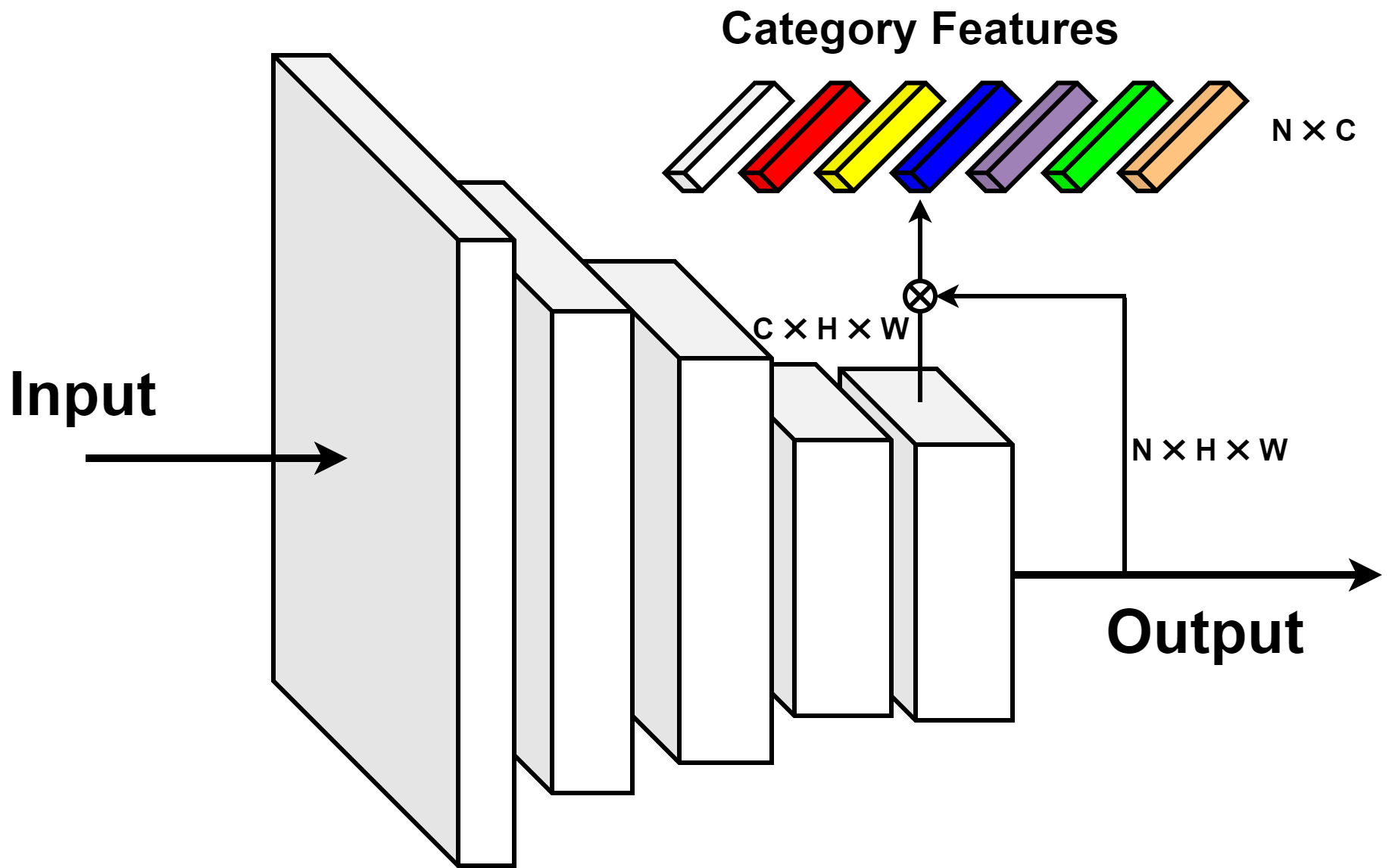}
	\caption{The framework of the Category Features Pooling (CFP) module.}
	\label{fig2}
\end{figure}

Specifically, we employ Deeplab v2 \cite{29} as the base segmentation model, where ResNet50 \cite{30} is used as the encoder $Enc$. According to \cite{13}, pixels in the same category cluster in the feature space, where the centroids of the features of each category can represent the category feature distributions. Thus, we propose the CFP module to extract the category features from the input image $X$  by combining the coarse outputs and the deep features. As shown in Fig. 2, we first extract the deep features $Enc(X)\in R^{{C}\times{H}\times{W}}$ and the coarse output ${Y^{'}}\in R^{{N}\times{H}\times{W}}$ from a given input image $X$, where $C$ and $N$ indicates the number of channels and the number of categories, $H$ and $W$ represent the height and width. Afterwards, the category features $f\in R^{{N}\times{C}}$ is calculated as:
\begin{equation}\label{CFP1}
	f = \delta(f_1, f_2)
\end{equation}
where
\begin{equation}\label{CFP2}
	\delta(f_1, f_2) = \frac{1}{H\times{W}}\sum_{k}^{H\times{w}}Y_k^{'}\times{Enc(X)}
\end{equation}

Likewise, the category features $f$ extracted by CFP are also used as the category centroids of the features of each category as a representative of the feature distribution. However, it is worth noting that different from the methods in \cite{13,14,15}, our $f$ are calculated in only one batch of images instead of all the images in the domain, which produces enormous time savings and an enhancement of efficiency. Also, we do not use the labels $Y$ to rectify the category features $f$ as in \cite{13,14,15}, since the coarse outputs $Y^{'}$ have been already supervised by the reliable labels $Y$.

\begin{figure*}
	\centering
	\includegraphics[width=1\linewidth]{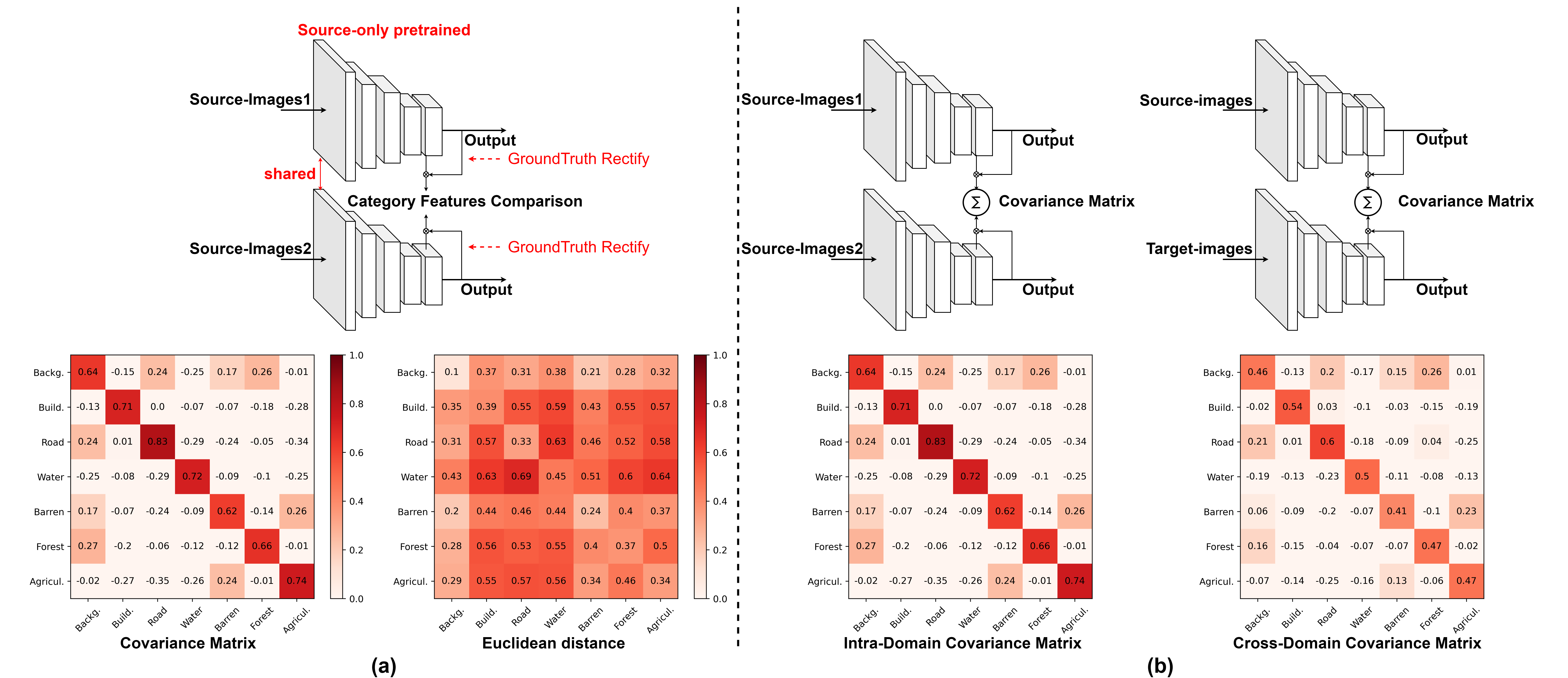}
	\caption{An explainable experiment for the proposed Covariance Regularization (CR). (a) is the comparison between our proposed Covariance matrix and the Euclidean distance used in \cite{13,14,15}. (b) further shows the discrepancy of category features from different domain, which is represented by our proposed Covariance matrix. It is worth noting that, the shared source-only pretrained network is used in both (a) and (b). The Ground-Truth Rectify is not used in the actual training as described in Section III A.}
	\label{fig3}
\end{figure*}

\subsection{Covariance Regularization (CR)}
\subsubsection{Motivations of CR}
To align the category features, we hope that the intra-category features can be closer and the inter-category features can be further apart. Given two category features $f_1$ and $f_2$ extracted by CFP module, we propose a Covariance Regularization (CR) method to align the features, which is defined as:
\begin{equation}\label{CR1}
	Corr(f_1, f_2) = \frac{E[({f_1}-\mu_{f_1})({f_2}-\mu_{f_2})]}{\sigma_{f_1}\sigma_{f_2}}
\end{equation}
where ${\sigma}_{f_1}$ and ${\sigma}_{f_2}$ are the average values of $f_1$ and $f_2$, ${\mu}_{f_1}$ and ${\mu}_{f_2}$ represent the variance of $f_1$ and $f_2$, respectively. $E$ represents the Expectation function. $Corr(f_1,f_2)$ builds a ${N}\times{N}$ Covariance matrix to represent the relationship between different category features. $Corr(f_1, f_2)$ is also known as the linear correlation coefficient in the field of mathematics, which is a Mahalanobis distance, instead of the Euclidean distance using in the aforementioned methods \cite{13, 14, 15}. 

To further illustrate our motivations of CR, we conduct an explainable experiment for the proposed CR as shown in Fig. 3. In Fig. 3(a), we divide two groups of images from source domain and input them through a source-only pretrained network. Then the category features comparisons are conducted for the $f_1$ and $f_2$ extracted by the CFP modules. It can be seen in Fig. 3(a) that, compared with the Euclidean distance using in \cite{13, 14, 15}, our proposed CR can represent the relationship between different category features much more explicitly. In addition, as shown in Fig. 3(b), we input the source images and the target images into the network to calculate the covariance matrix. We can see in Fig. 3(b) that the discrepancy between the source domain and the target domain can be clearly represented by the covariance matrix. With a covariance matrix, we hope the diagonal elements close to 1 and the off-diagonal elements less than 0, thus the CR is formulated as:
\begin{equation}\label{CR2}
\begin{aligned}
L_{CR}(f_1, f_2)=\frac{1}{N\times{N}}\sum_{i}^{N} \sum_{j}^{N} log(A_{ij}(Corr(f_1, f_2)))
\end{aligned}
\end{equation}
where
\begin{equation}\label{CR3}
A_{ij}(Corr)=\begin{cases}
Corr&i=j\\
max(1-Corr,\epsilon)&i\neq{j}\\
\end{cases}
\end{equation}
where $\epsilon$ is a small value to avoid logarithm with zero. 

We summarize the novelty of CR from three aspects. First, as shown in Fig. 3, our proposed CR is more robust when calculating the distance between category features. Since the large intra-category variance in RSIs will mislead the calculation of category features, Euclidean distance used in \cite{13,14,15} cannot represent the category distance distinctly. But according to Eq. (6), our CR is normalized with the variance $\sigma_f$, thus alleviate the impact of the intra-category variance. In addition, with the extremely imbalanced and inconsistent category distributions from two domains, the calculation of Euclidean distance almost depends on the category that with larger proportion, which cannot regularize the category features effectively. However, as a kind of Mahalanobis distance, our CR is also scale-invariant and not affected by dimensions, which means the category with small proportion can also impact the calculation of distance heavily. Thus, for RSIs with larger intra-category variance and inconsistent category distributions, our proposed CR can better align the category features between two domains than the existing category alignment methods.

\begin{figure*}
	\centering
	\includegraphics[width=1\linewidth]{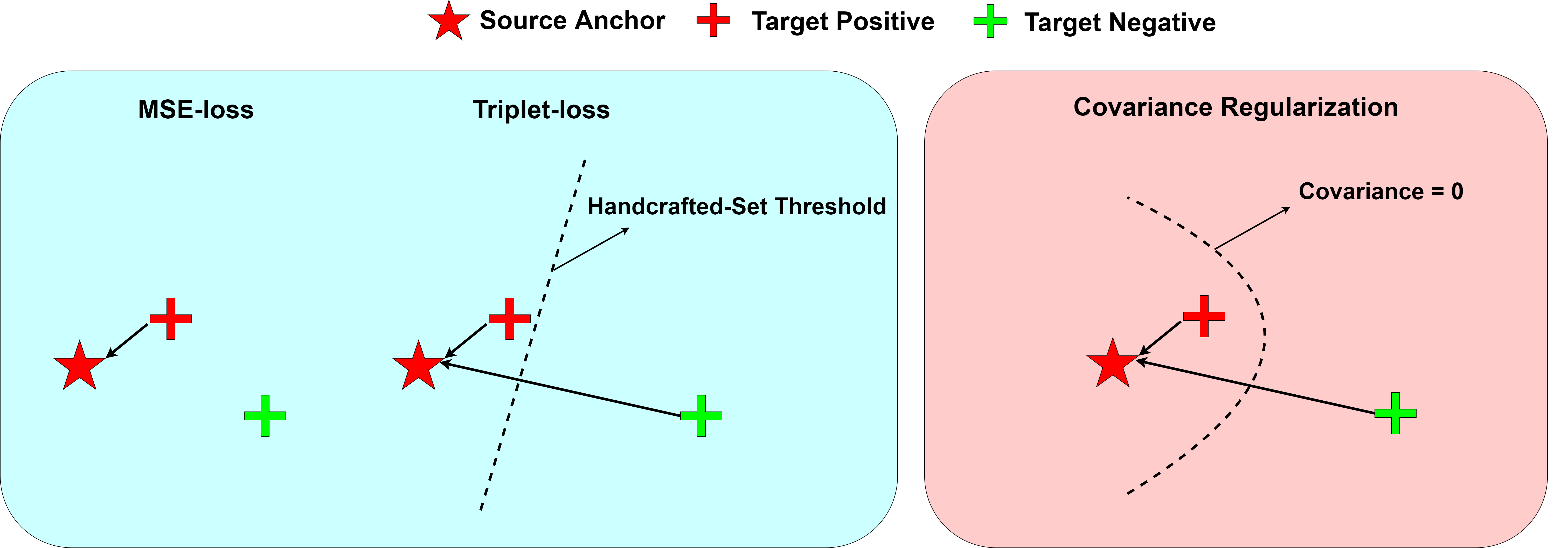}
	\caption{Comparisons between our proposed CR and the existing category alignment methods.}
	\label{fig4}
\end{figure*}

Second, unlike the Euclidean distance that only depends on the value difference, our CR concerns more about the relationship between different dimensions of the vectors. With high-dimension category features extracted, the existing methods \cite{13,14,15} utilize the Euclidean distance to impose a hard constraint to regularize the category features, but ignore the relationship between different dimensions. Instead, our CR encourages the model to extract linear-correlated or linear-independent features. For example, for the same categories, the CR enforces the same dimensions of features to be stimulated in the model, while invariant dimensions of features are stimulated for different categories. Thus, our CR can align the category features from the dimension aspects, instead of the unstably calculated value difference.

Finally, our proposed CR can adapt the UDA model more effectively compared with the existing category alignment methods. As shown in Fig. 4, our CR has outstanding advantages compared with the methods in \cite{13,14,15}. First, the MSE used in \cite{13,14} does not align different category features.Although the Triplet proposed in \cite{15} can regularize different category features, it is very difficult to learn the decision boundary when adjusting the relative magnitude of intra-category and inter-category simultaneously, which always requires a complex handcrafted threshold setting according to Eq. (3). However, our CR can not only align different category features, but also save the step to set the threshold, since the $corvariance=0$ is a native threshold. With $corvariance>0$, the category features are linear-correlated, while for $corvariance\leq{0}$ they are linear-independent or negative-correlated. Thus, the $corvariance=0$ is the best decision boundary for aligning intra-category and inter-category features, which does not require any other handcrafted settings. 

\subsubsection{Intra-domain CR and Cross-domain CR}

\begin{figure*}
	\centering
	\includegraphics[width=1\linewidth]{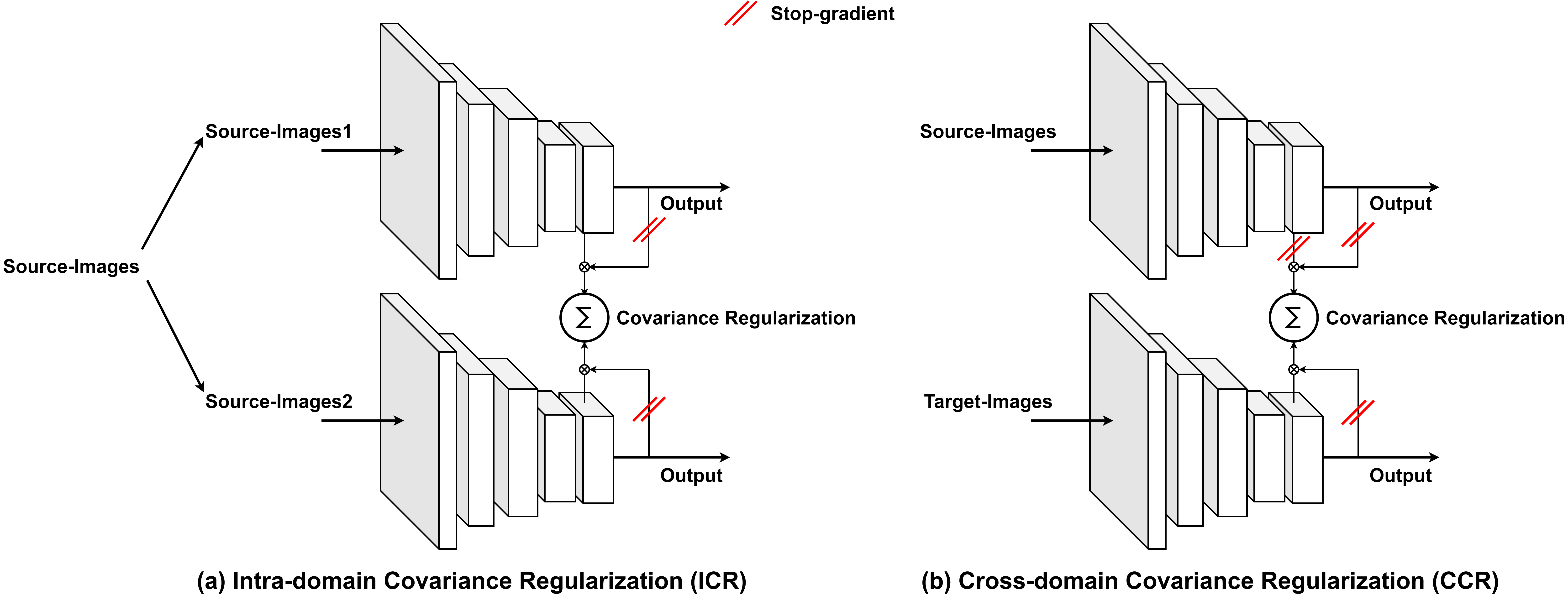}
	\caption{The architectures of the proposed (a) ICR and (b) CCR.}
	\label{fig5}
\end{figure*}

Intra-domain covariance regularization (ICR) and Cross-domain covariance regularization (CCR). As shown in Fig. 5(a), for ICR, given a batch of images, we first divide them into two groups randomly, then we perform CR to regularize the two group of category features. Inspired by \cite{33}, in ICR, we stop the gradient for outputs to optimize the learning of deep representations only, since the outputs are with reliable supervision from labels. It is worth noting that ICR is only employed in the source domain, since the coarse outputs in the source domain are supervised by reliable labels, which are much more accurate. The regularization loss $L_{ICR}$ is denoted as:
\begin{equation}\label{ICR}
	L_{ICR} = L_{CR}(f_s^1, f_s^2)
\end{equation}
where $f_s^1$ and $f_s^2$ are the category features from different groups in the source domain, respectively.

CCR is further employed to align the category features between different domains. While in CCR, category features from the source domain and outputs from both domains are fixed, the category features captured from target images are optimized only. Since the source domain are with more correct supervision, we assign them as credible reference, we can align the target features to the source features. Similarly, the $L_{CCR}$ is defined as:
\begin{equation}\label{CCR}
	L_{CCR} = L_{CR}(f_s, f_t)
\end{equation}
where $f_s$ and $f_t$ are the category features from the source domain and the target domain, respectively.

With the ICR and the CCR, we encourage the intra-category features closer and the inter-category features further separate in both of two domains, which adapts the proposed DCA model more progressively.

\begin{algorithm}
\caption{Stagewise training the DCA model}\label{algorithm}
\KwData{training dataset: $(X_s, Y_s, X_t)$, maximum stages: $K$, maximum iterations $I$}
\KwResult{Model $M_K$ and $(Y_s^{'}, Y_t^{'})$}

\While{$i \leq I$}
{
\If{$k = 0$}
{{Pretraining: $M_0$ $\leftarrow$ $(X_s, Y_s)$}\;
{$ICR$: $L_{ICR}$ $\leftarrow$ $(X_s, Y_s^{'})$ according to Eq.(9)}\;
}
\For {$k \leftarrow 1$ to $K$}
{{Generate $Y_t^{p}$ $\leftarrow$ $M_{k-1}$}\;
{$ICR$: $L_{ICR}$ $\leftarrow$ $(X_s, Y_s^{'})$ according to Eq.(9)}\;
{$CCR$: $L_{CCR}$ $\leftarrow$ $(X_s, Y_s^{'}, X_t, Y_t^{'})$ according to Eq.(10)}\;
{Training $M_k$ on ${(X_s, Y_s, X_t, Y_t^{p}, L_{ICR}, L_{CCR})}$ according to Eq.(14)} 
}
{$M_{k-1}$ $\leftarrow$ $M_k$}
}
\end{algorithm}

\subsection{Training of DCA}

Similar to the aforementioned ST methods, we train the DCA model with a stagewise procedure. The training procedure is proposed to avoid the accumulation of error-prone pseudo-labels generated in ST, which will produce incorrect supervision signals, leading to more erroneous pseudo-labels iteratively and trap the network to a local minimum with poor performance eventually. The procedure is shown in Algorithm 1. First, we pretrain the segmentation model \cite{29} on the source domain. A standard cross-entropy loss function $CE(Y^{'},Y)$ is used to supervise as:
\begin{equation}\label{CE1}
L_{CE}^s=CE(Y_s^{'}, Y_s)
\end{equation}
where
\begin{equation}\label{CE2}
\begin{aligned}
&CE(Y^{'}, Y)\\
 &= - \frac{1}{H\times{W}} \sum_{k}^{H\times{W}} \sum_{i}^{N} [Y_{ki} log(Y^{'}_{ki}) + (1 - Y_{ki}) log(1 - Y^{'}_{ki})]
\end{aligned}
\end{equation}

Meanwhile, the ICR is used to align the category features in the source domain. At the beginning of each stage, we generate or update the pseudo labels $Y_t^{p}$ to supervise the target predictions $Y_t^{'}$ as:
\begin{equation}\label{CE3}
L_{CE}^t=CE(Y_t^{'}, Y_t^{p})
\end{equation}

Next, we trained the DCA with $L_{CE}^s$, $L_{CE}^t$ and the proposed $L_{ICR}$ and $L_{CCR}$ for several stages:
\begin{equation}\label{TOTAL}
L=L_{CE}^s+ L_{CE}^t+L_{ICR}+L_{CCR}
\end{equation}

\section{Results OF Experiments}

To verify the effectiveness of our proposed method, we conducted extensive experiments on the advanced UDA RSI segmentation dataset LoveDA \cite{1}. In this section, we first introduce the dataset and the implementation details. Then, we perform detailed extensive ablation experiments on the dataset. Finally, we report our results on both Rural-to-Urban and Urban-to-Rural scenarios, including the visual results from different methods.

\subsection{Datasets}

The LoveDA dataset contains 5987 high-spatial-resolution images ($1024\times{1024}$ pixels) from three different cities. Compared to the existing datasets, the LoveDA dataset encompasses two domains (urban and rural), which focuses on improving the generalization capability of model from different urban and rural scenes. For Rural-to-Urban task, there are 1366 source images (rural) and 677 target images (urban) for training, and 820 target images (urban) for testing. As for Urban-to-Rural task, 1156 source images (urban) and 992 target images (rural) are used for training, while 976 target images (rural) for testing. The detailed data splits are described in \cite{1}.

\begin{table*}
	\setlength{\abovecaptionskip}{0.pt}
	\setlength{\belowcaptionskip}{-0.em}
	\centering
	\footnotesize
	\caption{Performance comparisons with baseline methods and category alignment methods}
\begin{threeparttable}
	\begin{tabular}{c|c|ccccccc|c}
		\hline
		\hline 
		\textbf{Domain}  &\textbf{Method} &\multicolumn{7}{|c|}{\textbf{IoU (\%)}} &\textbf{mIoU (\%)}\\
		\cline{3-9}
		 & &\textbf{Background} &\textbf{Building} &\textbf{Road} &\textbf{Water} &\textbf{Barren} &\textbf{Forest} &\textbf{Agriculture}\\
		 \hline
		&Source-only &42.45 &26.76 &25.08 &70.63 &12.98 &17.97 &26.83 &31.86 \\
		&Baseline &39.76 &43.78	 &33.91 &69.65	&8.69 &41.71 &22.67	 &37.17\\
		Rural-To-Urban &+MSE\cite{13}	&44.79	&42.12	&34.97 &79.97	&16.06	&31.44	&36.59	&40.85\\
		&+Triplet\cite{15} &40.69 	&43.78 &37.09 &80.76 &16.27 &20.78 &32.85	 &38.89\\
		\cline{2-10}
		&\textbf{DCA} &\textbf{45.82} &\textbf{49.60} &\textbf{51.65} &\textbf{80.88} &\textbf{16.70} &\textbf{42.93} &\textbf{36.92} &\textbf{46.36}\\
		\hline
		&Source-only &25.27	 &44.01	&22.64 &54.30	&6.97 &31.00 &39.93	 &32.02 \\
		&Baseline &26.72 &45.87	 &25.71 &58.46 &7.36 &38.73 &45.82	&35.52\\
		Urban-To-Rural &+MSE\cite{13}	&26.49 &46.28 &34.19 &60.45	 &1.15 &38.03 &54.65 &37.60\\
		&+Triplet\cite{15} &27.65 &45.12 &25.03 &58.68 &11.22 &37.93 &50.62 &36.61 \\
		\cline{2-10}
		&\textbf{DCA} &\textbf{36.38}	&\textbf{55.89} &\textbf{40.46} &\textbf{62.03} &\textbf{22.01}	&\textbf{38.92}	&\textbf{60.52}	&\textbf{45.17}\\
		\hline
		\hline
	\end{tabular}%
	\begin{tablenotes}
	\footnotesize               
        \item[1] Note: \textbf{+MSE} and \textbf{+Triplet} means employ MSE based alignment \cite{13} and Triplet based alignment \cite{15} with the Baseline, respectively. And Baseline is the common ST strategy described in Section IV.C(1). 
    \end{tablenotes}            
    \end{threeparttable}
	\label{table:Align}%
\end{table*}%

\begin{figure*}
	\centering
	\includegraphics[width=1\linewidth]{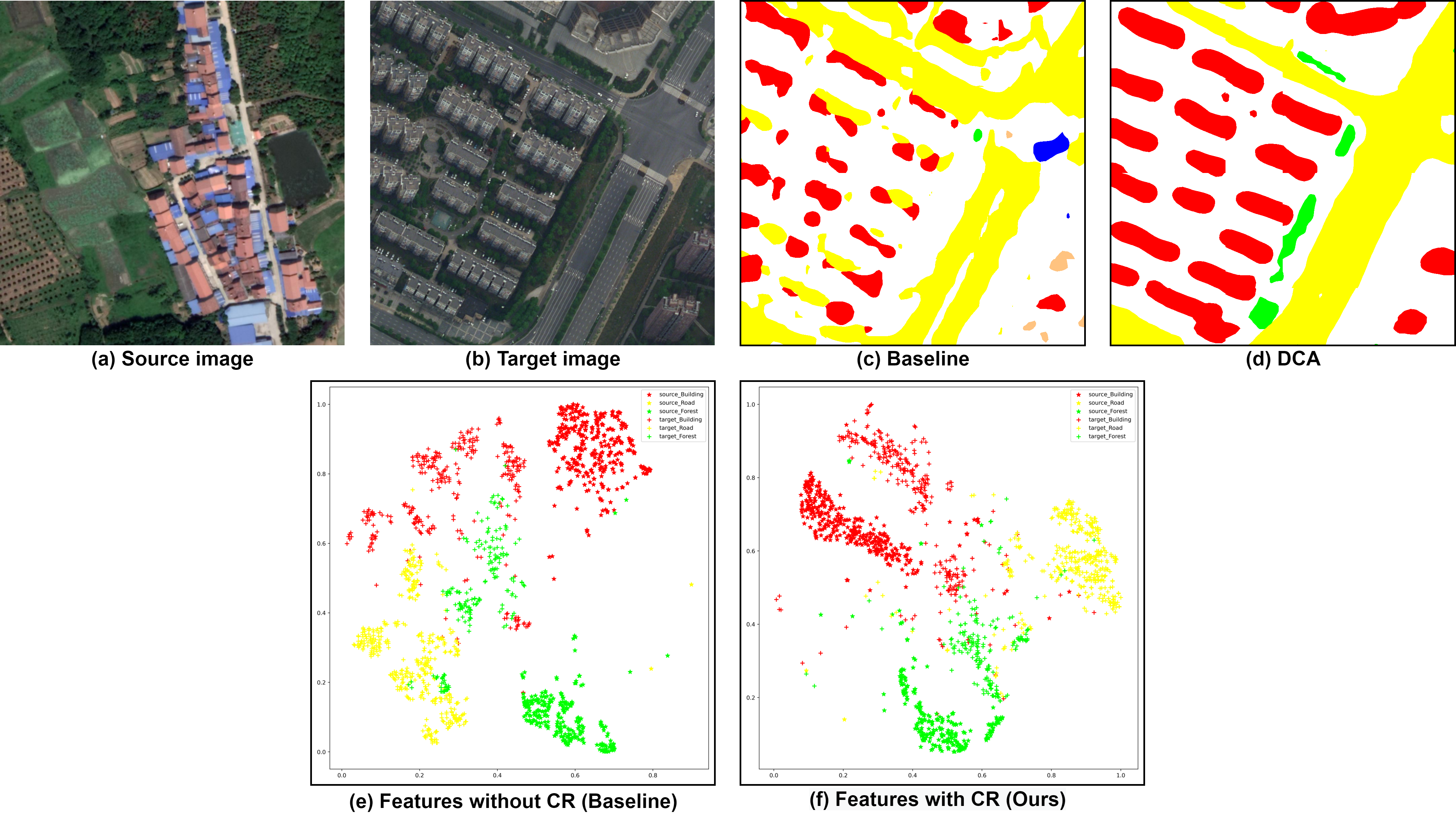}
	\caption{A contrastive analysis of the proposed DCA and the baseline method on the Rural-to-Urban experiment. (a) is a source image from rural domain, (b) is a target image from urban domain. (c) and (d) are the segmentation results of the baseline method and our DCA method, respectively. We then map the high-dimension features of (c) and (d) to a 2D space with t-SNE \cite{40}, shown in (e) and (f).}
	\label{fig6}
\end{figure*}

\subsection{Experimental Settings}
\subsubsection{Experimental Environment }
All experiments were processed on a server computer with an NVIDIA A100 GPU.  The implementation of the framework was based on the open-source toolbox Pytorch. Each experiment is conducted for five times and the average results are reported.

\subsubsection{Implementation Details}
We use the intersection over union (IoU) to report the semantic segmentation accuracy. With respect to the IoU for each class, the mIoU represents the mean of the IoUs over all the categories.

All the methods adopt the same feature extractor and discriminator, Specifically, DeepLabV2 \cite{29} with ResNet50 \cite{30} was utilized as the extractor, and the discriminator was constructed by fully convolutional layers \cite{6}. The backbones used in all the networks were pre-trained on ImageNet \cite{31}. According to the experiments in \cite{1}, during the training, we used the Stochastic Gradient Descent (SGD) optimizer with a momentum of 0.9 and a weight decay of 0.0005. The learning rate was initially set to 0.01, and a poly schedule with power 0.9 was applied. For the data augmentation, $512\times{512}$ patches were randomly cropped from the raw images ($1024\times{1024}$ pixels) for training, with random mirroring and rotation. While in the testing part, we perform slide-predictions for the raw images, in order to alleviate the resolution's impact on the segmentation results. In the stagewise training Algorithm 1, the maximum stages K was set to 5 and the maximum iterations I was set to 10000 for both two experiments.

\subsection{Ablation Study for DCA}

\subsubsection{Comparisons with Category Alignment methods}
To validate the efficiency of our DCA, we first compare our DCA method with the source-only method and the baseline method. Here, we define the baseline method as using the common ST strategy, which simply pretrained the model in the source domain and generated the pseudo labels to supervise the target images according to Eq. (13) directly without any other prior. Next, we further conducted a fair comparison between the proposed CR and other category alignment methods, \emph{i.e.}, MSE based alignment in \cite{13}, Triplet based alignment in \cite{15}. As for Triplet alignment method, we tried several times to find the best $\alpha$ according to Eq. (3) and finally set to 0.5. The detailed results are shown in Table I.

As shown in Table I, the ST baseline can promote the generalization of source-only pretrained model by a large margin. In addition, the MSE alignment method brings improvements in mIoU, which illustrates the effectiveness of category alignment methods. However, the Triplet alignment method is not effective to enhance the accuracy. The reason may be that the training of triplet is complex and not available for UDA RSI segmentation. In addition, our proposed DCA can achieve a remarkable improvement compared with other methods. In Rural-to-Urban experiment, our DCA yields a 9.19\% improvement in mIoU and 9.65\% improvement in mIoU for Urban-to-Rural experiment, which outperforms the other methods with a large margin. This demonstrates that our approach can greatly improve the generalization capability of model by aligning category features.

\begin{table*}
	\setlength{\abovecaptionskip}{0.pt}
	\setlength{\belowcaptionskip}{-0.em}
	\centering
	\footnotesize
	\caption{Performance comparisons with state-of-the-art UDA segmentation methods in rural-to-urban experiment}
	\begin{tabular}{c|c|ccccccc|c}
		\hline
		\hline 
		\textbf{Method}  &\textbf{Type} &\multicolumn{7}{|c|}{\textbf{IoU (\%)}} &\textbf{mIoU (\%)}\\
		\cline{3-9}
		 & &\textbf{Background} &\textbf{Building} &\textbf{Road} &\textbf{Water} &\textbf{Barren} &\textbf{Forest} &\textbf{Agriculture}\\
		 \hline
		Source-only &- &42.45 &26.76 &25.08 &70.63 &12.98 &17.97	 &26.83 &31.86 \\
		MCD\cite{41}	&- &43.60 &15.37 &11.98	 &79.07 &14.13 &33.08	&23.47	&31.53\\
		AdaptSeg\cite{6} &AT &42.35 &23.73 &15.61 &81.95	&13.62	&28.70	&22.05	&32.68\\
		FADA\cite{7}	&AT	&43.89	&12.62	&12.76 &80.37	&12.70	&32.76	&24.79	&31.41\\
		CLAN\cite{8}	&AT	&43.41	&25.42	&13.75 &79.25	&13.71	&30.44	&25.80	&33.11\\
		TransNorm\cite{9} &AT	&38.37	&5.04	&3.75	&80.83	&14.19	&33.99	&17.91 &27.73\\
		PyCDA\cite{10}	&ST	&38.04	&35.86	&45.51 &74.87	&7.71	&40.39	&11.39	&36.25\\
		CBST\cite{11}	&ST	&48.37	&46.10	&35.79 &80.05	&19.18	&29.69	&30.05	&41.32\\
		IAST\cite{12} 	&ST	&\textbf{48.57}	&31.51	&28.73	&\textbf{86.01}	&\textbf{20.29}	&31.77	&36.50	&40.48\\
		\hline
		\textbf{DCA}	&ST	&45.82	&\textbf{49.60}	&\textbf{51.65}	&80.88	&16.70	&\textbf{42.93}	&\textbf{36.92}	&\textbf{46.36}\\
		\hline
		\hline
	\end{tabular}%
	\label{table:Rural2Urban}%
\end{table*}%

\begin{table*}
	\setlength{\abovecaptionskip}{0.pt}
	\setlength{\belowcaptionskip}{-0.em}
	\centering
	\footnotesize
	\caption{Performance comparisons with state-of-the-art uda segmentation methods in urban-to-rural experiment}
	\begin{tabular}{c|c|ccccccc|c}
		\hline
		\hline 
		\textbf{Method}  &\textbf{Type} &\multicolumn{7}{|c|}{\textbf{IoU (\%)}} &\textbf{mIoU (\%)}\\
		\cline{3-9}
		 & &\textbf{Background} &\textbf{Building} &\textbf{Road} &\textbf{Water} &\textbf{Barren} &\textbf{Forest} &\textbf{Agriculture}\\
		 \hline
		Source-only &- &25.27	&44.01	&22.64	&54.30	&6.97	&31.00	&39.93	&32.02 \\
		MCD\cite{41}	&- &25.61	&44.27	&31.28	&44.78	&13.74	&33.83	&25.98	&31.36\\
		AdaptSeg\cite{6} &AT &26.89	 &40.53	&30.65	&50.09	&16.97	&32.51	&28.25	&32.27\\
		FADA\cite{7}	&AT	&24.39	&32.97	&25.61	&47.59	&15.34	&34.35	&20.29	&28.65\\
		CLAN\cite{8}	&AT	&22.93	&44.78	&25.99	&46.81	&10.54	&37.21	&24.45	&30.39\\
		TransNorm\cite{9} &AT	&19.39	&36.30	&22.04	&22.04	&\textbf{36.68}	&14.00	&40.62	&24.62\\
		PyCDA\cite{10}	&ST	&12.36	&38.11	&20.45	&57.16	&18.32	&36.71	&41.90	&32.14\\
		CBST\cite{11}	&ST	&25.06	&44.02	&23.79	&50.48	&8.33	&\textbf{39.16}	&49.65	&34.36\\
		IAST\cite{12} 	&ST	&29.97	&49.48	&28.29	&\textbf{64.49}	&2.13	&33.36	&\textbf{61.37}	&38.44\\
		\hline
		\textbf{DCA}	&ST	&\textbf{36.38}	&\textbf{55.89}	&\textbf{40.46}	&62.03	&22.01	&38.92	&60.52	&\textbf{45.17}\\
		\hline
		\hline
	\end{tabular}%
	\label{table:Urban2Rural}%
\end{table*}%

\begin{figure*}
	\centering
	\includegraphics[width=1\linewidth]{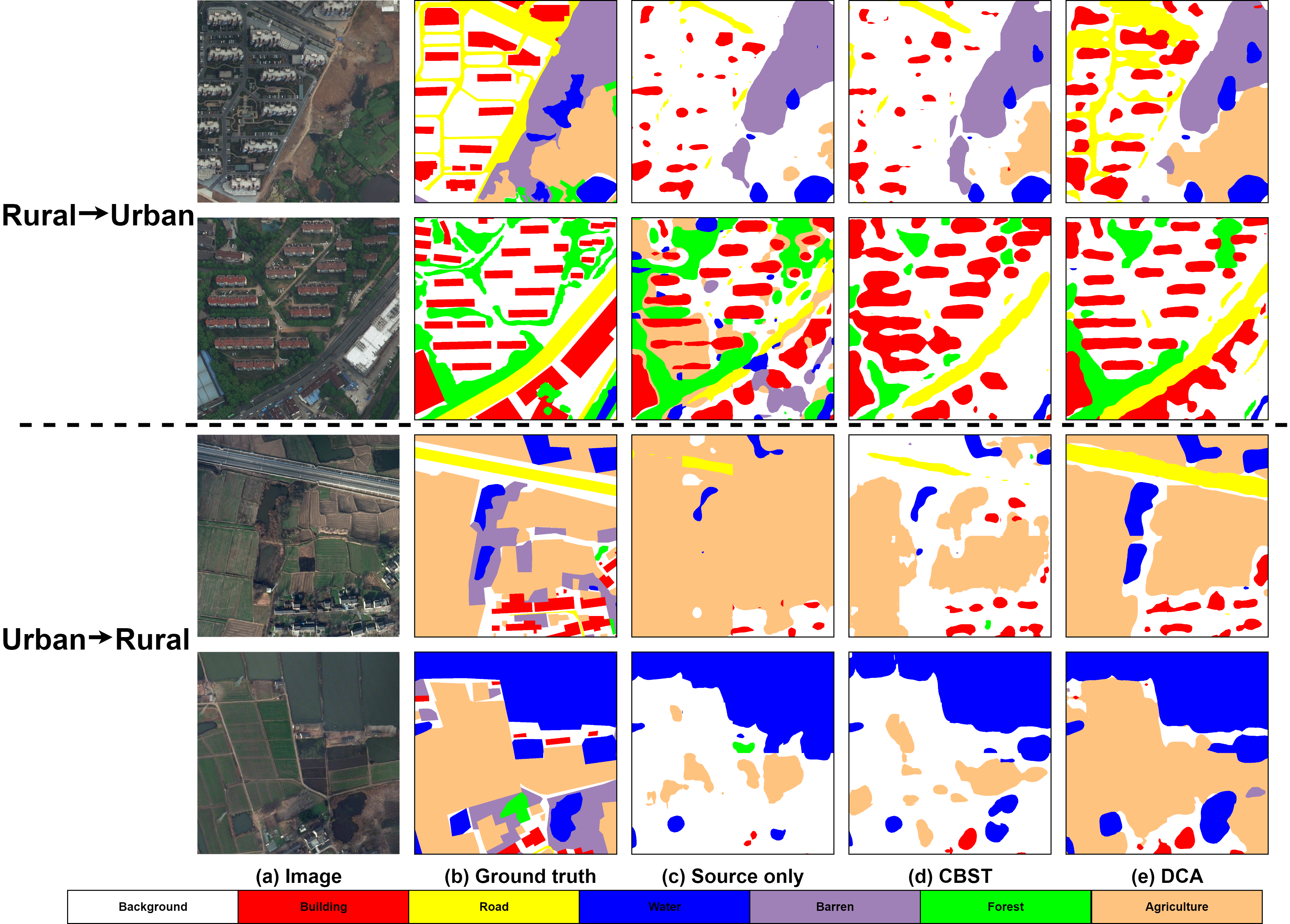}
	\caption{Qualitative comparisons between the proposed methods and other methods on the Val dataset.}
	\label{fig7}
\end{figure*}

\begin{figure*}
	\centering	\includegraphics[width=1\linewidth]{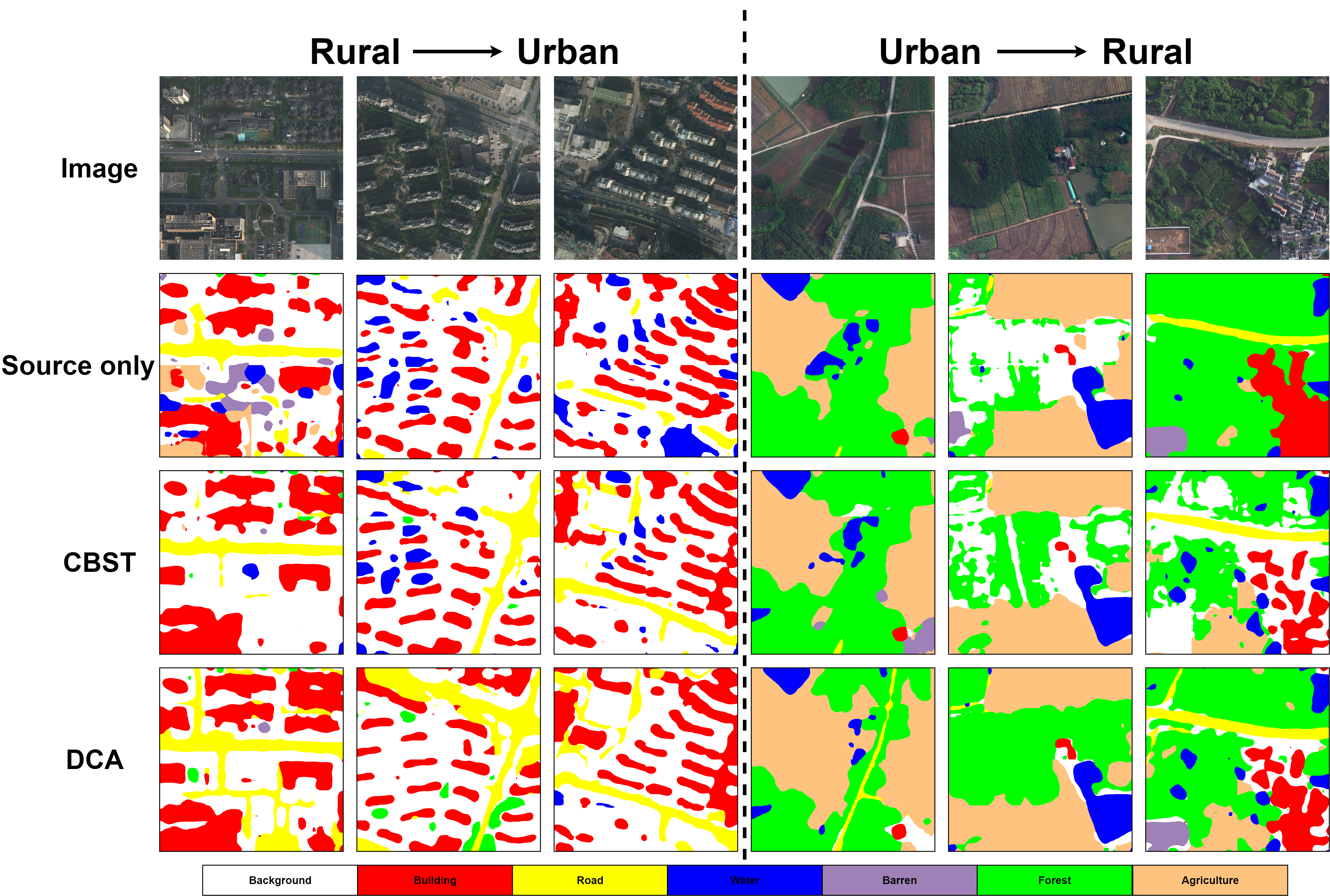}
	\caption{Qualitative comparisons between the proposed methods and other methods on the Test dataset.}
	\label{fig8}
\end{figure*}

\subsubsection{Visualization Results of Category Features Distributions}
To further illustrate the novelty of our proposed DCA, we visualize the deep category features distributions in latent space. To this end, given a source image and a target image with large domain discrepancy, we map their high-dimensional deep features to a 2D space with t-SNE \cite{40}, as shown in Fig. 6. We focus on the feature distributions between different categories from two domains. As shown in Fig. 6(e), for the baseline method, the learned features of buildings, roads, forests are separate due to the domain shift. It can be seen that without CR alignment, the baseline method cannot explicitly learn the shared representations of the same categories from different domains. Instead, we can observe in Fig. 6(f) that our proposed DCA can drive the intra-category features closer and inter-category features further apart, which build more accurate UDA segmentation results shown in Fig. 6(d). 

\subsubsection{Comparisons with UDA Segmentation methods}
The LoveDA dataset contains two scenarios: Rural-to-Urban and Urban-to-Rural. We carry out experiments for the both scenarios on the LoveDA dataset \cite{1} to evaluate the effectiveness of our DCA compared with other state-of-the-art UDA segmentation methods, as shown in Tables II and III.

It can be seen in Table II, in the Rural-to-Urban experiments, the DCA achieves the highest mIoU of 46.36\%, which outperforms the state-of-the-art method CBST by 5.04\%. Specifically, the DCA brings a great improvement of the IoU of buildings, roads and forests. However, for background and barren, the IoUs do not improve obviously. The reason may be that the features of these two categories are too complex to align. For the Urban-to-Rural experiments shown in Table III, DCA also achieves a mIoU of 45.17\% with a large gap compared with other methods. The improvements are mainly from buildings and roads, which means our DCA can effectively extract the shared features of these categories from the rural domain to urban domain. The results demonstrate the efficiency of our proposed DCA method. 

\subsubsection{Visualization Results}
The qualitative results are shown in Figs. 7 and 8. Since we do not have the corresponding Ground Truth on the Test dataset, we also display the results on the Val dataset in Fig. 7 in order to compare with the Ground-Truths. Fig. 8 shows the visualization results on the Test dataset.

It can be seen that our DCA can produce better UDA segmentation predictions than the source-only method and CBST \cite{11}. Specifically, in Rural-to-Urban experiments, our DCA predicts more complete constructions for the buildings and roads. As in the Urban-to-Rural experiments, the segmentation results of agriculture are more accurate. The visualization results can also demonstrate the effectiveness of our proposed framework.

\section{Conclusion}
In this paper, we presented a novel Deep Covariance Alignment (DCA) framework for unsupervised domain adaptive remote sensing image segmentation. Our proposed DCA used a novel Covariance Regularization (CR) to enforce the intra-category features to be closer and the inter-category features to be further separate, and thus explicitly extracted shared domain-invariant discriminative feature representations to enhance the ability of model generalization. Extensive experiments on the LoveDA dataset demonstrated the effectiveness and efficiency of the proposed DCA compared with other state-of-the-art methods.

In the future, we will explore the development of the quantitative experiments to analyze the visualization and interpretation of results performed by our DCA and compared to other approaches. Moreover, we will investigate the proficiency of the domain adaptive segmentation task in remote sensing images and consider other alignment approaches to further improve segmentation performance.

\ifCLASSOPTIONcaptionsoff
  \newpage
\fi

\bibliographystyle{unsrt}

\end{document}